\documentclass[9pt, conference]{IEEEtran}
\usepackage{graphicx}				
\usepackage{xcolor}
\usepackage{amsmath,epsfig,amsfonts,amssymb,epsfig,amsbsy}
\usepackage{mathrsfs}
\usepackage{wrapfig}
\usepackage{amssymb}

\usepackage{algorithmic,algorithm}

\newcommand{\Real}{\mathbb{R}}

\def \bP{\mathbf{P}}

\def \bX{\mathbf{X}}
\def \bY{\mathbf{Y}}

\def \bphi{\mathbf{\Phi}}

\def \br {\mathbf{r}}

\def \bGamma {\mathbf{\Gamma}}

\def \be {\mathbf{e}}

\def \bN{\mathbf{N}}
\def \bY{\mathbf{Y}}

\def \bZ{\mathbf{Z}}

\newcommand \bu{\mathbf{u}}

\def \b1 {\mathbf{1}}

\begin{document}
\title{Methods for Large Scale Hydraulic Fracture Monitoring}
\vspace{-5mm}
\author{ \IEEEauthorblockN{Gregory Ely and Shuchin Aeron}
\IEEEauthorblockA{Dept. of ECE, Tufts University, Medford, MA 02155\\
\textbf{Email}: gregory.ely@tufts.edu, shuchin@ece.tufts.edu}}

\maketitle

\begin{abstract}
In this paper we propose computationally efficient and robust methods for estimating the moment tensor and location of micro-seismic event(s) for large search volumes. Our contribution is two-fold. First, we propose a novel joint-complexity measure, namely the sum of nuclear norms which while imposing sparsity on the number of fractures (locations) over a large spatial volume, also captures the rank-1 nature of the induced wavefield pattern. This wavefield pattern is modeled as the outer-product of the source signature with the amplitude pattern across the receivers from a seismic source. A rank-1 factorization of the estimated wavefield pattern at each location can therefore be used to estimate the seismic moment tensor using the knowledge of the array geometry. In contrast to existing work this approach allows us to drop any other assumption on the source signature. Second, we exploit the recently proposed first-order incremental projection algorithms for a fast and efficient implementation of the resulting optimization problem and develop a hybrid stochastic \& deterministic algorithm which results in significant computational savings. 
\end{abstract}

\section{Introduction}
Seismic hydraulic fracture monitoring (HFM) can both mitigate many of the environmental risks and improve reservoir effectiveness by providing real time estimates of locations and orientations of induced fractures.  Determining the location of these microseimsic events remains challenging due to high levels of pumping noise, propagation of seismic waves through highly anisotropic shale, and the layered stratigraphy leading to complex wave propagation \cite{Eisner_TLE09}.
Classical techniques for localization involves de-noising of individual traces \cite{Liu_IGARSS09,Rodriguez_Geophy12} followed by estimating the arrival time of the events at each individual trace.  The angle of arrival of the incident array, or polarization, is achieved via Hodogram analysis \cite{Han_CREWES_2010_Thesis} or max-likelihood type estimation \cite{khadhraoui_SEG10}.  Once the angle and time arrival of the events has been estimated, the events are back-propagated using a forward model under known stratigraphy to determine the location \cite{khadhraoui_SEG10}. In contrast to these approaches which tend to separate the de-noising of the signal from the physical model, recently the problem of moment tensor estimation and source localization was considered in \cite{Sacchi_GJI12} for general sources and in \cite{Ely_IGARSS12, Ely_ICA2013} for isotropic sources which exploit sparsity in the number of microseimsic events in the volume to be monitored. This approach is shown to be more robust and can handle processing of multiple time overlapping events.

Our approach, although similar to the technique proposed in \cite{Sacchi_GJI12}, differs in that we do not use source waveform information from the Green's function and introduce a group low-rank penalization. Here we don't use the amplitude of the received waveform, but only the fact that the received signal across the seismometers is common across all seismometers with varying delays dictated by a known velocity model of the stratigraphy and the source receiver configuration.  Since we are not using any amplitude information, we usually have more error in estimation and require more receivers for localization. Nevertheless when the computation of Green's function is costly or accurate modeling of the stratigraphy is not available, our method can be employed. Furthermore, due to amplitude independent processing our methods can be extended to handle the anisotropic case using just the travel-time information for inversion \cite{Pratt_1992}.
\begin{figure}
\begin{center}
      \includegraphics[width = .34\textwidth]{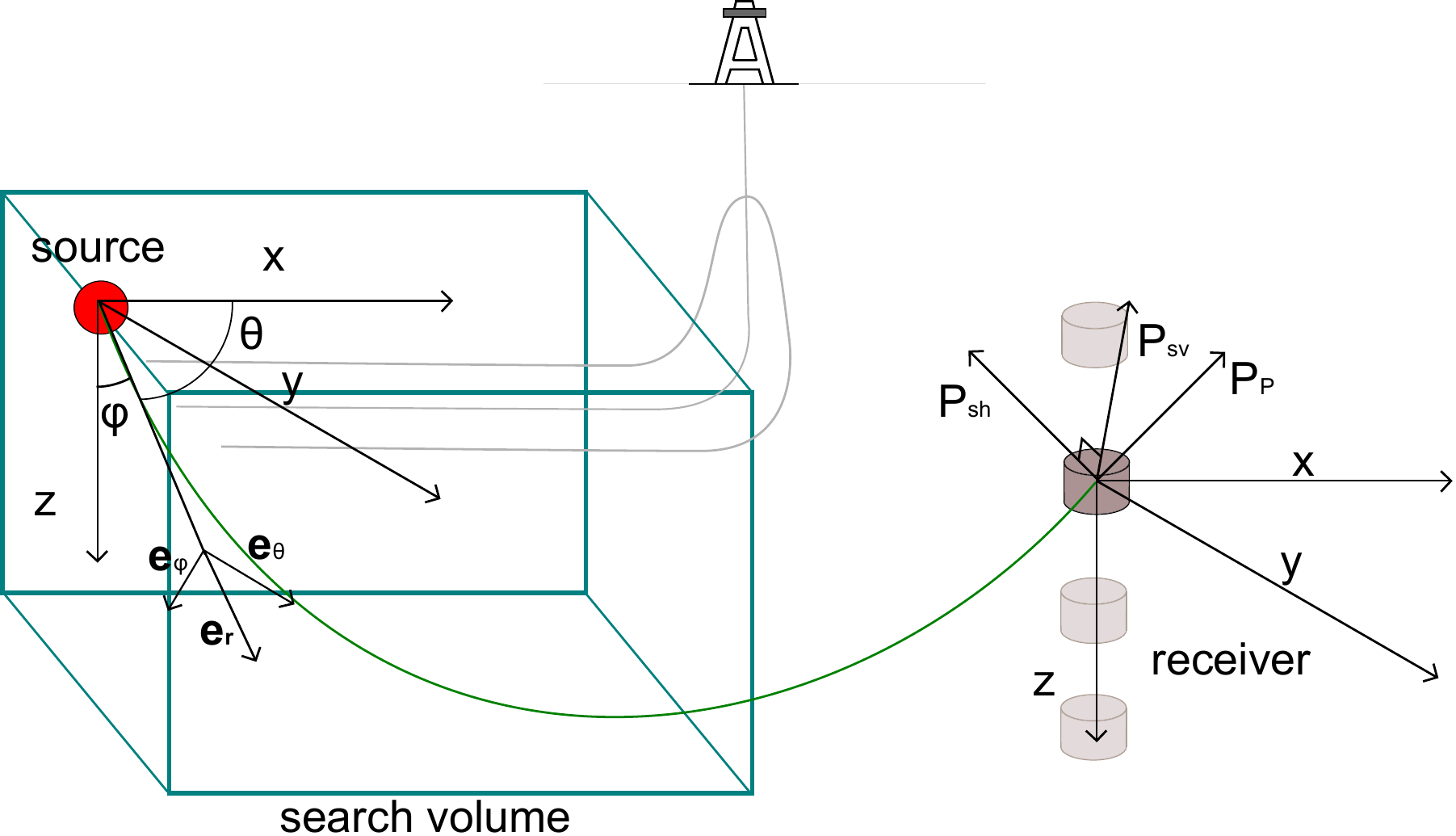}
      \end{center}
        \caption{This figure shows the geometry and coordinate system used throughout this paper.}
  \label{fig:setup}
\end{figure}
\vspace{-2mm}
\section{Physical model}
In this paper we focus on  propagation in isotropic media, although our approach can easily be extended to anisotropic and layered media. Figure~\ref{fig:setup} shows the physical setup in which a seismic event with a symmetric moment tensor $\mathbf{M} \in \Real^{3\times3}$ is recorded at a set of $J$ tri-axial seismometers indexed as $j = 1,2,...,J$ with locations $\br_{j}$ and $I$ denotes the location of the seismic event $l$. The seismometer record compressional wave denoted by $p$, and vertical and horizontal shear waves denoted by $sv$ and $sh$ respectively. Assuming that the volume changes over time does not change the geometry of the source, Equation~($\ref{eq:particle}$) describes the particle motion vector  $\bu_c(l,j,t)$  at the three axes of the seismometer $j$ as a function of time $t$.
\begin{align}
\label{eq:particle}
\bu_c(l,j,t) = \frac{R_c(\theta,\phi)}{4\pi d_{lj} \rho c^3} \,\,\mathbf{P}_{c}^{l j} \,\psi_c\left(t - \frac{d_{lj}}{v_c}\right)
\end{align}
where $d_{lj}$ is the radial distance from the source to receiver; $c \in \{p, sh, sv\}$ is the given wave type, and $\rho$ is the density, and $R_c$ is the radiation pattern which is a function of the moment tensor, the take off direction parameters $\theta_j, \phi_j$ with respect to the receiver $j$.  $\mathbf{P}_{c}^{l j}$ is the unit polarization vector for the wave $c$ at the receiver $j$. Up to a first order approximation \cite{madariaga_seismic_2007} we assume that $\psi_c(t) \approx \psi(t)$ for all the wave types and henceforth will be referred to as the source signal. Note that for non-anisotropic formations the compressional waves $\bP_{p}^{l j}$ aligns with the direction of ray propagation. The polarization vectors for the $sh$ and $sv$ correspond to the other mutually perpendicular directions.
The radiation pattern depends on the moment tensor $\mathbf{M}$ and is related to the take off direction at the source with respect to the receiver $j$ defined as the radial unit vector $\mathbf{e}_{r_j}$ relative to the source as determined by $(\theta_j , \phi_j)$, see Figure~\ref{fig:setup}. Likewise we denote the unit vectors $\be_{\theta_j}$ and $\be_{\phi_j}$  to be the radial coordinate system orthogonal to radial unit vector. The radiation pattern for a compressional source $R_p(\theta_j,\phi_j)$ is then given by,
\begin{align}
\begin{split}
R_p(\theta_j,\phi_j) = \mathbf{e}_{r_j}^T  \mathbf{M}  \mathbf{e}_{r_j} 
\end{split}
\begin{split}
\mathbf{M} = 
\left[
\begin{array}{ccc}
  M_{xx} & M_{xy} & M_{xz} \\
  M_{xy} & M_{yy} & M_{yz} \\
  M_{xz} & M_{yz} & M_{zz} \\
\end{array}
\right]
\end{split}
\end{align}
The radiation energy at a receiver can then be simplified and described as the inner product of the vectorized compressional unit vector product, $\textbf{e}_{p_j}$, and the vectorized moment tensor $\textbf{m}$; where $(\cdot)^T$ denotes the transpose operation.
\begin{align}
\begin{array}{l}
R_p(\theta_j,\phi_j) = {\textbf{e}_{p_j}^{T}} \mathbf{m}; \,\,\,\,\mathbf{m} = [M_{xx}, M_{xy}, M_{xz}, M_{yy}, M_{yz}, M_{zz}]^{T}\\
\\
\textbf{e}_{p_j}^{T} = [e_{r_{jx}}^2, 2e_{r_{jx}} e_{r_{jy}},  2e_{r_{jx}}e_{r_{jz}},  e_{r_{jy}}^2,  2e_{r_{jy}}e_{r_{jz}},  e_{r_{jz}}^2]^T
\end{array}
\end{align} The above expression can then be extended to construct a vector of radiation pattern $\mathbf{a}_p \in \Real^{J}$ across the $J$ receivers, with take off angles of ($\theta_j$ and $\phi_j$) corresponding to compressional unit vectors $\textbf{e}_{p_j}$, given by $\mathbf{a}_{p}= \mathbf{E}_p \mathbf{m}$ where $\mathbf{E}_p = [\be_{p_1},\be_{p_2},...,\be_{p_J}]^T$.

Similarly we have $ \mathbf{a}_{sh} = \mathbf{E}_{sh} \mathbf{m} $ and $ \mathbf{a}_{sv} = \mathbf{E}_{sv} \mathbf{m} $. Therefore we can write the radiation pattern across $J$ receivers for the three wave types as the product of an augmented matrix with the vectorized moment tensor.
\begin{align}
\label{eq:amp_model}
\mathbf{a} =
\left[
\begin{array}{l}
    \mathbf{a}_p  \\
    \mathbf{a}_{sh}   \\
    \mathbf{a}_{sv}  \\
  \end{array}
\right]
=
\underset{\mathbf{E}}{\underbrace{\left[
\begin{array}{l}
    \mathbf{E_p}  \\
    \mathbf{E_{sh}}   \\
    \mathbf{E_{sv}}  \\
  \end{array}
\right] }}\mathbf{m}
\end{align}
Thus the radiation pattern across the receivers, $\textbf{a}$, can then be described as the product of the $\textbf{E}$ matrix, which depends on the location of the event and the configuration of the array, and the vectorized moment tensor, derives solely from the geometry of the fault.  Under the above model for seismic source and wave propagation, given the noisy data at the tri-axial seismometers, the problem is to estimate the event location and the associated moment tensor.  This separability will be exploited in our dictionary construction to better recover the location and characteristics of the source. 

\begin{figure}[ht]
\centering \makebox[0in]{
\begin{tabular}{c}

      \includegraphics[width = .34\textwidth]{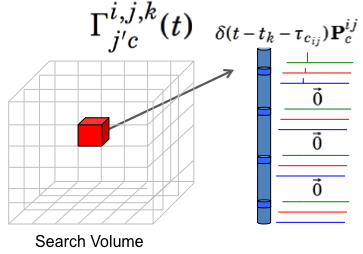}
      \end{tabular}}
  \caption{This figure shows an example propagator from a single seismic source for our dictionary construction.}
  \label{fig:dict}
\end{figure}
\vspace{-3mm}
\section{Dictionary Construction}
\vspace{-1 mm}
Our approach relies on the construction of a suitable representation of the data acquired at the receiver array under which seismic events can be \emph{compactly represented}. We then exploit this compactness to robustly estimate the event location \& moment tensor.

Under the assumption that the search volume  $I$ can be discretized into  $n_V$ locations indexed by $l = l_1, l_2,...,l_i,..,l_{n_V}$, we construct an over complete dictionary of space time propagators $\bGamma_{c}^{i,j,k}$.  Where $\bGamma_{c}^{i,j,k}$ describes the noiseless data at the \emph{single} receiver, $j$, as excited by an \emph{impulsive} hypothetical seismic event $i$ at location $l_i$ and time $t_k$ of wave type $c$ ($p$,$sh$ or $sv$).  Figure \ref{fig:dict}  shows a pictorial representation of a single propagator.
\begin{align}
\label{eq:propagator}
\mathbf{\Gamma}_{j' c}^{i,j,k}(t) =
\begin{cases}
        \delta(t- t_ k - \tau_{{c}_{ij}} ) \, \bP_{c}^{ij}  & \text{if } j' = j \\
        \vec{0} & \text{if } j' \neq j ,
\end{cases}
\end{align}Note that $\tau_{c_{ij}} = \frac{d_{l_i j}}{v_c}$ is the time delay and $\mathbf{\Gamma}_{j' c}^{i,j,k} \in \Real^{ |\mathbb{T}_{r}|\times J \times 3}$.   We then construct a dictionary $\Phi$ of propagators for all locations, source time indices, wave types, and receiver indices, where each row of the dictionary represents a vectorized propagator,
\begin{align}
\Phi= [ \bGamma_{c}^{i,1,k}(:), \bGamma_{c}^{i,2,k}(:), \hdots, \bGamma_{c}^{n_v,J,k}(:) ] 
\end{align}
where $(:)$ denotes the MATLAB colon operator which vectorizes the given matrix starting with the first dimension.   Because the dictionary covers all possible locations, receiver responses, time support of the signal, and wave types, an observed seismic signal $\bY$ in the presence of Gaussian noise $\bN$ can be written as the superposition of numerous propagators,
\begin{align}
\bY = \Phi \bX(:) +\bN
\end{align}
where $\bX$ is the coefficient tensor of size ${3\cdot J \times |\mathbb{T}_{s}| \times n_{V}}$ and each of there dimensions correspond to $1^{st}$ wave type receiver index, $2^{nd}$ source time index, and $3^{rd}$ location index as shown in Figure \ref{fig:dict}.  

\begin{figure}[ht]
\centering \makebox[0in]{
\begin{tabular}{c}
      \includegraphics[width = .35\textwidth, height = 1.3in]{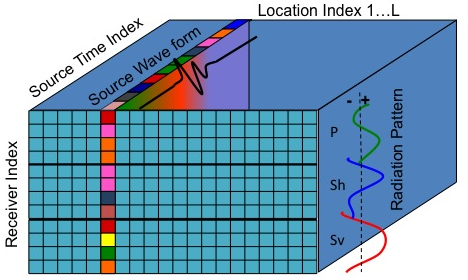}
      \end{tabular}}
  \caption{This figure shows the block sparsity.The \textbf{\emph{lateral}} slice of the dictionary coefficients corresponding to the correct location is a rank-1 outer product of the source signal and the amplitude pattern.}
  \label{fig:dict2}
\end{figure}

Therefore, for the case of  a single seismic event with some radiation pattern and arbitrary source signal, the coefficient tensor $\bX$ will be block sparse along the lateral slice, with the only non-zero slice corresponding to location index $l$, see Figure \ref{fig:dict2}. Furthermore, the observed source signal will be common across all of receiver indices of the dictionary with its amplitude modulated by the radiation pattern.  Therefore, the dictionary elements will not only be block sparse, but the non-zero slice can be written as, $\boldsymbol{\psi} \,\mathbf{a}^{T}$, i.e. a rank-1 outer-product of the radiation pattern at the source wave signal.  This notion can be extended to the case of  (small number of) multiple events where $\bX$ will have now have a few non-zero rank-1 slices. This is the key observation which we exploit in this paper as discussed below.
\vspace{-3mm}
\section{Algorithms for large scale HFM}
\vspace{-1mm}
Under the above formulation and the assumption that for a given recorded signal contains only a few seismic events, we exploit the block-sparse/low-rank, structure of $\bX$ for robust HFM by proposing a complexity penalized recovery methods.

In \cite{Ely_ICA2013} we proposed the following group sparse penalized optimization set-up \cite{Tropp2006SPa} for robust localization of the seismic event,
\begin{align}
\label{eq:minl12}
\hat{\bX} = \underset{\bX}{\arg\min}\,\,||\bY(:) - \bphi \bX(:)||_{2}^2 + \lambda \sum_{i = 1}^{n_V} ||\bX(:,:,i)||_{2}
\end{align}
where $||\bX(:,:,i)||_{2}$ denotes the $\ell_2$ (Frobenius) norm of the $i$-th slice, $\lambda$ is a sparse tuning factor that controls the group sparseness of $\bX$, i.e. the number of non-zero slices, versus the residual error.  The parameter $\lambda$ is chosen depending on the noise level and the anticipated number of events. The location estimate is then given by selecting the slices with the largest Frobenius norm above some threshold.  

In order to exploit the block low-rank structure of the dictionary coefficients, we propose the following group nuclear norm penalized optimization set-up,
\begin{align}
\label{eq:minl1nuc}
\hat{\bX} =  \underset{\bX}{\arg\min}\,\, ||\bY(:) - \bphi \bX(:)||_{2}^2 + \lambda \sum_{i = 1}^{n_V} ||\bX(:,:,i)||_{*}
\end{align}
where $||\bX(:,:,i)||_{*}$ represents the nuclear norm, i.e. the sum of the singular values of the $i$-th slice.

\textbf{Iterative Algorithms}: To solve either of the optimization problems given in Equations (\ref{eq:minl12}) \& (\ref{eq:minl1nuc}) we implemented three different forms of first order algorithms, Iterative Shrinkage (ISTA), Fast Iterative Shrinkage (FISTA) \cite{Beck:2009gh} and stochastic gradient descent with incremental proximal methods \cite{Bertsekas:2011ww}. ISTA being the simplest to implement is given by two operations: a gradient descent step, and a shrinkage operation like so,
\begin{align}
\label{eq:prox}
\bX^{k+1} = \mathbf{prox}_{\frac{\lambda}{\alpha}}(\bX^{k} - \frac{1}{\alpha}\Phi^T(\Phi \bX^k - \bY)  )
\end{align}
where $\alpha$ is the step size and $\mathbf{prox}_\tau(z)$ is the shrinkage operator for one of the two norms.   For the group sparse minimization the shrinkage operation is given by,
\begin{align}
\mathbf{prox l12}_\tau(\mathbf{Z}) = \underset{\bX}{\arg \min}\,\,\frac{1}{2}||\bX-\bZ||_{2}^2 + \tau \sum_{i = 1}^{n_V} ||\bZ(:,:,i)||_{2}
\end{align}
and for the group low-rank the prox-operator is equivalent to a shrinkage on the singular values of each of the lateral slices $\bX(:,:,i)$ of $\bX$.
\begin{align}
& \mathbf{prox nuc}_\tau(\bZ(:,:,i)) \nonumber \\
& = \underset{\bX}{\arg\min}\,\,\frac{1}{2}||\bX(:,:,i)-\bZ(:,:,i)||_{2}^2 + \tau ||\bZ(:,:,i)||_{*}
\end{align}
Iterative shrinkage can be increased in speed with minimal overhead by adding an interpolation term resulting in the FISTA algorithm.
\begin{align}
\begin{array}{l}
\bZ = \bX^k + \frac{k-1}{k+2}(\bX^k - \bX^{k-1}) \\
\\
\bX^{k+1} = \mathbf{prox}_{\frac{\lambda}{\alpha}}(\bZ - \frac{1}{\alpha}\Phi^T(\Phi \bZ - \bY)  )
\end{array}
\end{align}
The resulting FISTA algorithm achieves convergence in $O(1/k^2)$ iterations vs $O(1/k)$ for ISTA.  In the case of the group low-rank penalization the proximal iteration can be  expensive to calculate given the large number of SVDs that need to be computed.  

\textbf{Incremental Proximal Method}: For large scale problems it becomes computationally infeasible to calculate the full proximal iteration.  As the problem scales, the gradient also becomes more expensive to calculate at each iteration.  Stochastic gradient descent with incremental proximal iterations can alleviate the computation burden by descending along a random subset of the full gradient and only applying the proximal shrinkage to  a few random slices at each iteration \cite{Bertsekas:2011ww}.  
Given that the penalty term in Equation~(\ref{eq:minl1nuc}) can be written as the sum of nuclear norms, the calculation of the shrinkage operation can be significantly reduced by only applying the shrinkage to a few slices per iteration.  In this application of stochastic gradient descent our forward operator $\Phi$ is sparse resulting in negligible difference in computational cost if the full or partial gradient is calculated.  Therefore we can apply the full gradient at each iteration $k$ and the proximal minimization operation to a subset $J \subset\{1,2,...,n_V\}$ of size $m_k$, Algorithm \ref{alg:stochNuc}. We implemented two forms of the incremental proximal method: `\textbf{dynamic}' where the number of subsets on which the shrinkage operation is applied is increased linearly per iteration, i.e. $m_{k+1} = \min(m_k + \beta ,n_V)$ where $\beta$ is a positive integer, or `\textbf{fixed}' where $m_k$ remains constant, say $m_0$, for all iterations. 
 
 \begin{algorithm}
 \caption{Incremental Proximal: solves (Eq. \ref{eq:minl1nuc})}
 \begin{algorithmic}
 \label{alg:stochNuc}
 \STATE $\bX = \textbf{0}$, $m_0$ //Initialize variables.
 \STATE Option = ``dynamic" or ``fixed"
 \WHILE{Not Converged}
 	\STATE if (Option == dynamic) \hspace{5mm}$m_k = \min(m_{k-1} + \beta,n_v)$, \\
 	       else if (Option == fixed) \hspace{3mm} $m_k = m_0$
 	\STATE $J = randperm(m_k,n_V)$ // \hspace{2mm} generate group index
 	\STATE $\textbf{Z} = \textbf{X} - \frac{1}{\rho}\Phi^T(\Phi\textbf{X}-\textbf{Y})$ // \hspace{2mm} Gradient calculation
 	\STATE $\textbf{X} = \textbf{Z}$ 
 	\STATE //Apply shrinkage operator only to indices $\in J$
 	\FOR{$j \in J$}
 	\STATE ${\textbf{X}}(:,:,{j}) =  \mathbf{prox nuc}_{\frac{\lambda}{\rho}}[\textbf{Z}(:,:,{j})]$ 
 	\ENDFOR
 \ENDWHILE
 \end{algorithmic}
 \end{algorithm}

\textbf{Moment tensor estimation from the slices}: 
To estimate the moment tensor we use the predicted event location source-receiver array configuration to construct the matrix $\mathbf{E}$. Then using the estimate of the radiation pattern $\hat{\mathbf{a}}$ from the left singular vector of the active slice we construct the inverse problem $\hat{\mathbf{a}} = \mathbf{E} \mathbf{m}$ and apply Tikhonov regularization to mitigate the ill-conditioning of the $\mathbf{E}$ operator.  The moment tensor vector $\mathbf{m}$ is estimated via,
\begin{align}
\hat{\mathbf{m}}=((\mathbf{E}^T\mathbf{E}+\lambda_{m} \mathbf{I})^{-1}\mathbf{E}^T)\hat{\mathbf{a}}
\label{eq:moment}
\end{align}
where $\lambda_m$ is again tuned using some estimates on the uncertainty in estimation of $\mathbf{a}$ and according to the amount of ill-conditioning of $\mathbf{E}$.
\vspace{-5mm}
\section{Experiments}
\begin{figure}
\centering \makebox[0in]{
\begin{tabular}{c}
      \includegraphics[width = .24\textwidth]{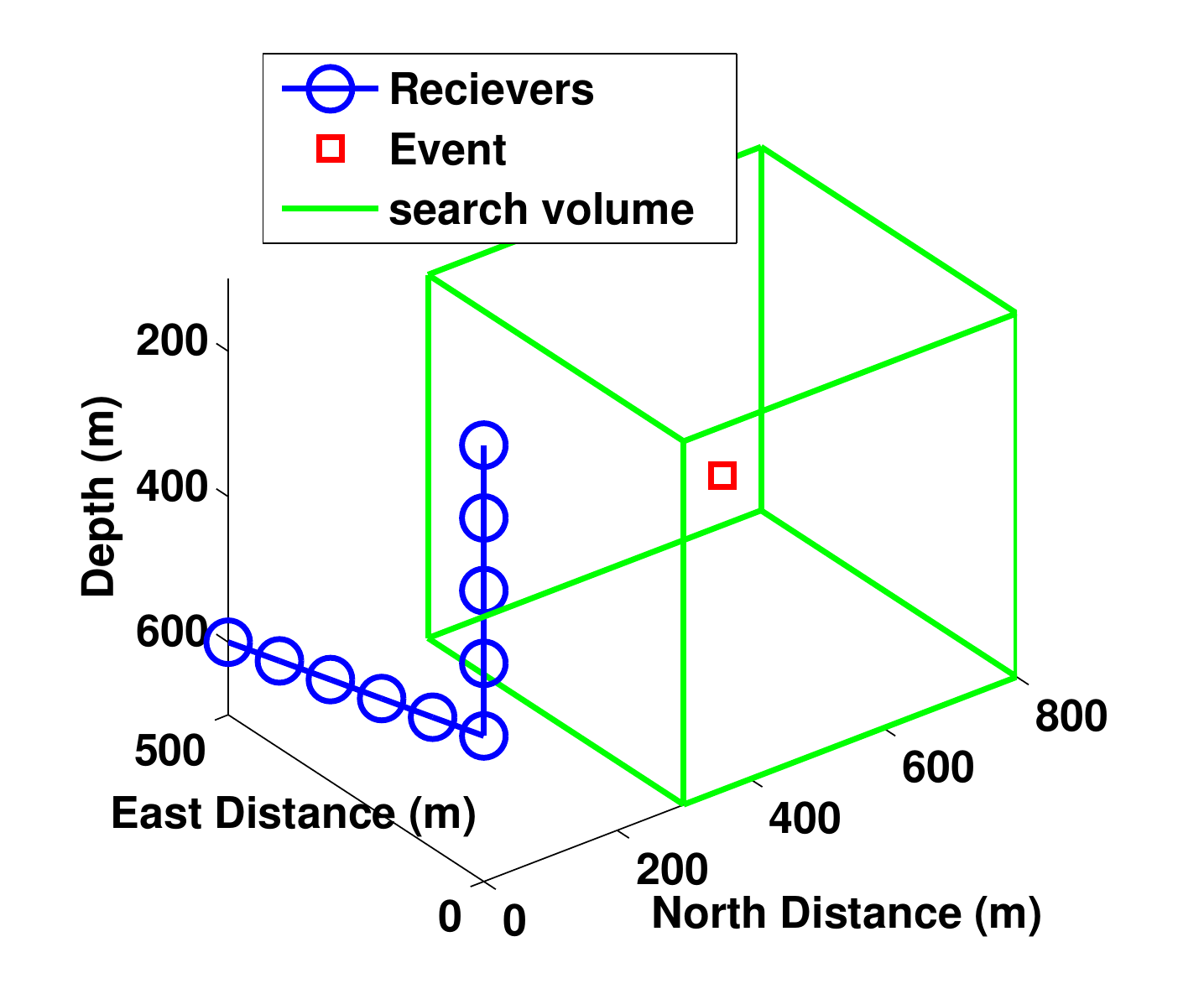}
      \includegraphics[width = .24\textwidth]{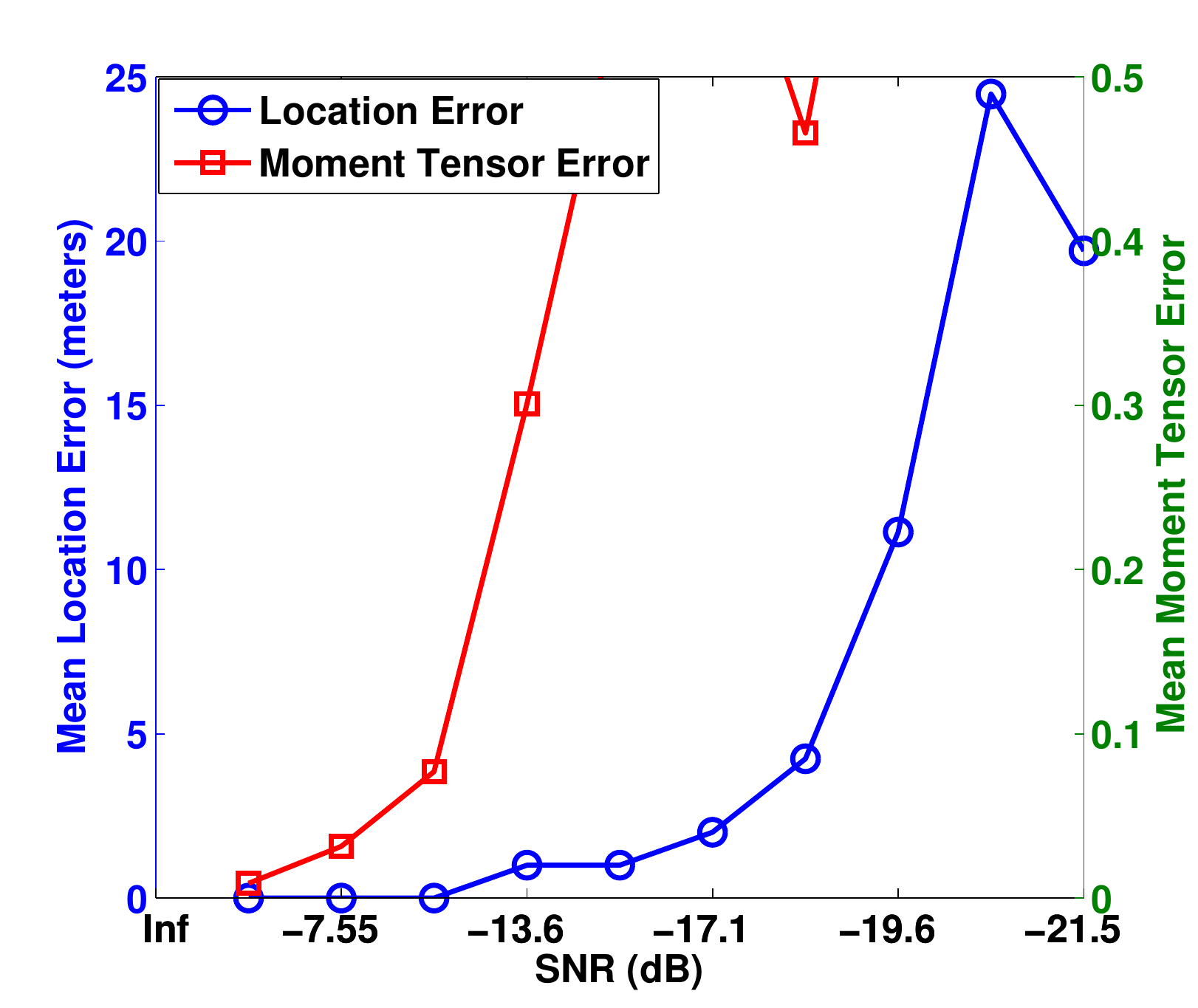}
      \end{tabular}}
        \caption{ \textbf{Left: }This figure shows the setup for the deviated well and the search volume used in the experiment section. \textbf{Right:} This figure show location and moment tensor error as a function of SNR.}
    \label{fig:noise}
\end{figure}

\vspace{-2mm}
In order to test the effectiveness of the proposed algorithm we simulated an array of 10 seismometers equally spaced within a deviated well consisting of a 500 meter vertical and 500 meter horizontal section dipping at 20 degrees relative to horizontal and aligned with the Y axes, as shown in Figure \ref{fig:noise} left. For the sake of simplicity, the earth is considered to be isotropic with compressional velocity of 1500 and shear velocity of 1100 meters per second.  A search volume of $500\times500\times500$ meters was placed perpendicular to well centered at (500, 300, 500) meters with varying resolution depending on the specific experiment conducted.  

\textbf{Performance in Noise}: In order to determine the effectiveness of the algorithm in the presence of noise, an explosive event with a shear component event was generated in the center of the search volume was generated with an increased grid resolution of 5 meters in the presence of various noise levels varying from 0 to -21 dB.  The minimization operation given by Equation \ref{eq:minl1nuc} was then applied to resulting simulations with a $\lambda$ of .9 and the location index with the largest nuclear norm was taken to be the location of the event.  Equation \ref{eq:moment} with a $\lambda$ of  .01 was then used to invert for the moment tensor.  This process was repeated 15 times for each noise level and the mean location error and RMS error in the estimate of the moment tensor vector are shown in Figure \ref{fig:noise} right.

\textbf{Algorithm Speed}: In order to test the speed of the three algorithms, the search volume was configured with a coarse spatial resolution of 25 meters and the same event as in the previous section was generated in Gaussian noise with a resulting SNR of -18 dB.  The three first order algorithms, ISTA, FISTA, and Incremental Proximal, were then applied to the group low-rank minimization problem, Equation~(\ref{eq:minl1nuc}), with a $\lambda$ of .9 and step size of $.5*10^3$.    Given that the search volume consisted of 9261 locations each iteration of both FISTA and ISTA would involve the computation of 9261 SVDs of matrices of size $N_t$ x $3N_r$.  In the case of incremental proximal method the number of SVD's taken per iteration could be set to 1 to 9261 per iterations.  Furthermore, because the forward operator for this problem is sparse and thus fast to compute, the entire full gradient was calculated at each iteration. The two variants of the incremental proximal algorithm, namely the ``fixed" and  ``dynamic" were implemented with $m_0 = 100$ and $\beta= 5$. 

\begin{figure}
\begin{center}
      \includegraphics[width = .35\textwidth]{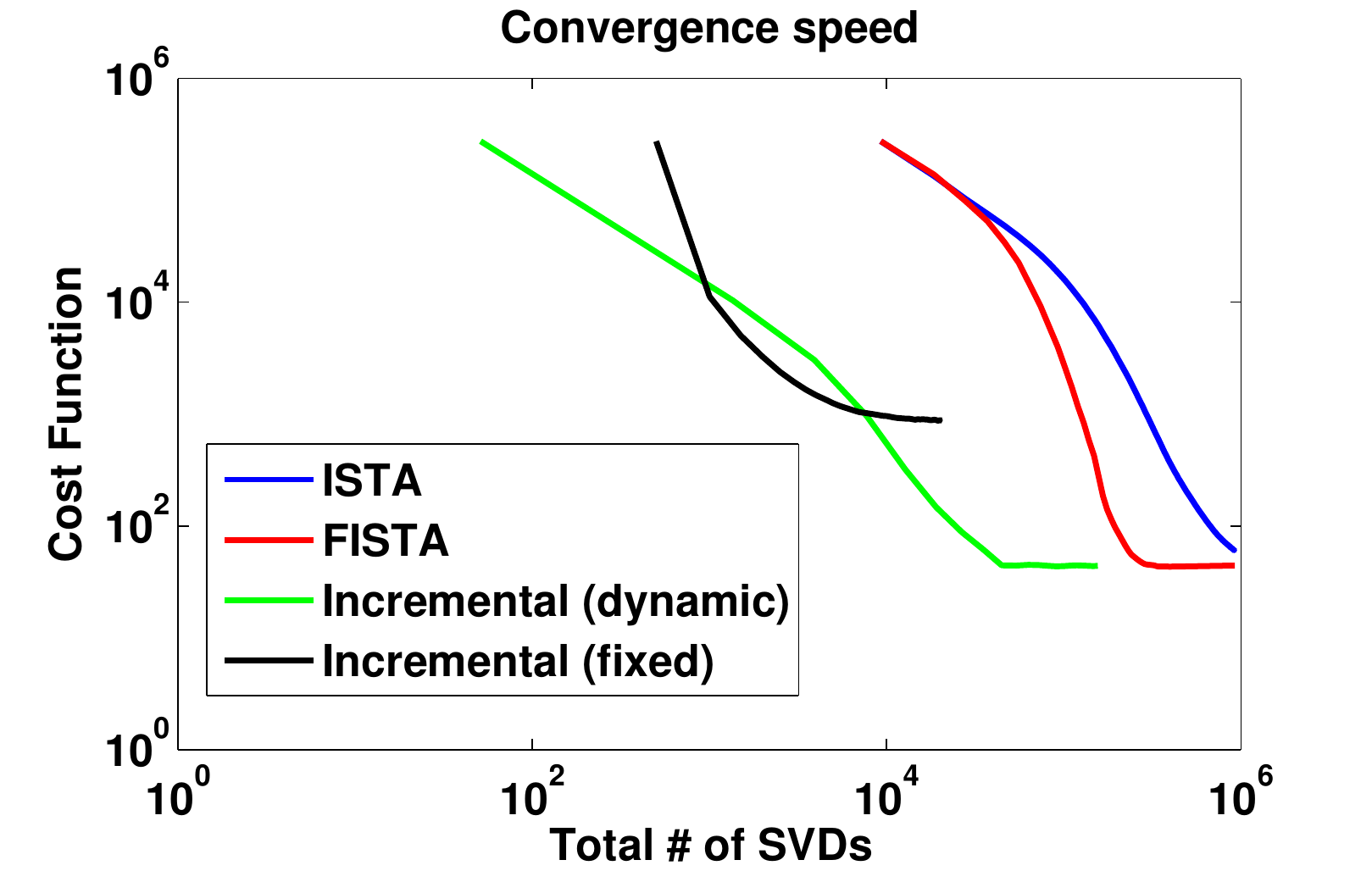}
      \end{center}
        \caption{This figure shows the convergence of the objective function, Equation~(\ref{eq:minl1nuc}), as a function of number of SVDs computed.}
  \label{fig:speed}
\end{figure}

Figure \ref{fig:speed} shows the convergence results for the various algorithms showing the cost function, Equation~(\ref{eq:minl1nuc}), as a function of total number of SVDs computed.  As expected FISTA outperforms ISTA and the incremental fixed method results in early convergence.  The incremental method with an increasing number of SVDs converges to the global minima in drastically fewer SVDs than either FISTA or ISTA.

\begin{figure}
\begin{center}
      \includegraphics[width = .45\textwidth]{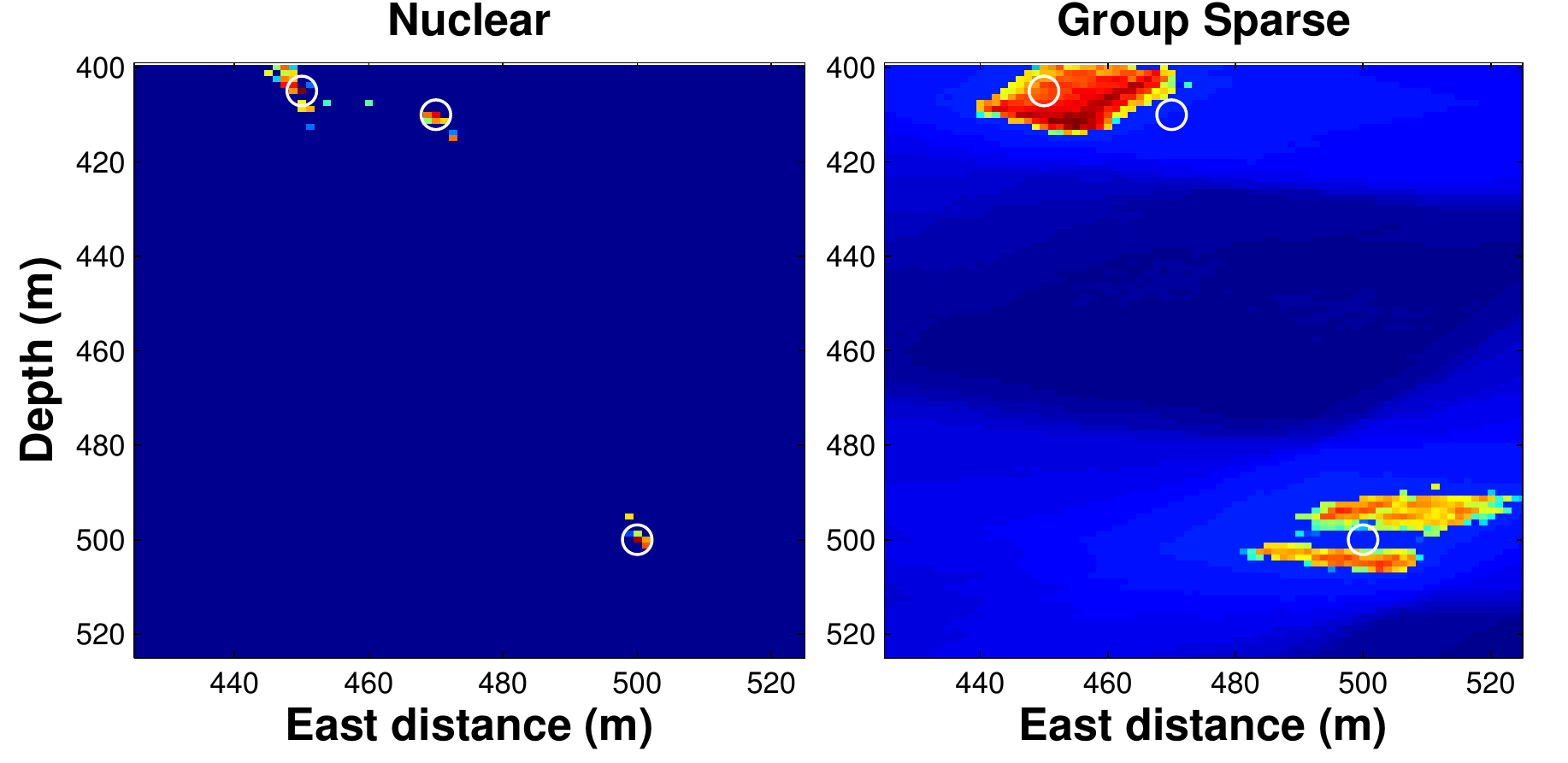}
      \end{center}
        \caption{Performance in source localization for the group $\ell_2$ sparse vs group nuclear sparse minimization algorithms.  Image intensities are shown on a log scale. The location of the true events are shown with white circles.}
  \label{fig:mult}
\end{figure}

\textbf{Multiple Events}: 
In order to test the algorithms ability to distinguish multiple events, three seismic events with varying moment tensors were generated in moderate noise within the search volume with a spatial resolution of 1.25 meters all with the same Y location such that the three events occupied a plane perpendicular to the X and Z axes.   Both the group $\ell_2$ sparse and group nuclear minimization operations were applied to the simulation with a $\lambda$ of .9.  Figure \ref{fig:mult} shows resulting nuclear and Frobenius norms along the X-Z plane after the minimization operation have been applied.  In the case of the nuclear norm minimization, three distinct events are visible falling precisely on the location of true events.  However for the group sparse penalization, the location of the two near incident events are impossible to separate and the outlying event's location is imprecise.

\textbf{Acknowledgment}:
This material is based upon work supported by the National Science Foundation Graduate Research Fellowship under Grant No. DGE-0806676.
\vspace{-4mm}
\bibliographystyle{IEEEbib}
\bibliography{camsap,Opt_bib}

\begin{thebibliography}{10}

\bibitem{Eisner_TLE09}
Leo Eisner and Peter~M. Duncan,
\newblock ``Uncertainties in passive seismic monitoring,''
\newblock {\em The Leading Edge 28}, vol. 28, pp. 648--655, 2009.

\bibitem{Liu_IGARSS09}
Qiuhua Liu, S.~Bose, H.-P. Valero, R.G. Shenoy, and A.~Ounadjela,
\newblock ``Detecting small amplitude signal and transit times in high noise:
  Application to hydraulic fracture monitoring,''
\newblock in {\em IEEE Geoscience and Remote Sensing Symposium}, 2009.

\bibitem{Rodriguez_Geophy12}
Ismael Vera~Rodriguez, David Bonar, and Mauricio Sacchi,
\newblock ``Microseismic data denoising using a 3c group sparsity constrained
  time-frequency transform,''
\newblock {\em Geophysics}, vol. 77, no. 2, pp. V21--V29, 2012.

\bibitem{Han_CREWES_2010_Thesis}
Lejia Han,
\newblock {\em Microseismic Monitoring and Hypocenter Location},
\newblock Ph.D. thesis, Department of Geoscience, Calgary, Alberta, Canada,
  2010.

\bibitem{khadhraoui_SEG10}
Bassem Khadhraoui, David Leslie, Julian Drew, and Rob Jones,
\newblock ``Real-time detection and localization of microseismic events,''
\newblock {\em SEG Technical Program Expanded Abstracts}, vol. 29, no. 1, pp.
  2146--2150, 2010.

\bibitem{Sacchi_GJI12}
I.~V. Rodriguez, M.~Sacchi, and Y.~J. Gu,
\newblock ``Simultaneous recovery of origin time, hypocentre location and
  seismic moment tensor using sparse representation theory,''
\newblock {\em Geophysical Journal International}, 2012.

\bibitem{Ely_IGARSS12}
G.~Ely and S.~Aeron,
\newblock ``Robust hydraulic fracture monitoring (hfm) of multiple time
  overlapping events using a generalized discrete radon transform,''
\newblock in {\em Geoscience and Remote Sensing Symposium (IGARSS), 2012 IEEE
  International}, july 2012, pp. 622 --625.

\bibitem{Ely_ICA2013}
Gregory Ely and Shuchin Aeron,
\newblock ``Complexity penalized hydraulic fracture localization and moment
  tensor estimation under limited model information,''
\newblock in {\em Proceedings of Meetings on Acoustics (POMA)}, Montreal,
  Canada, 2013, vol.~19, p. 045051, Acoustical Society of America.

\bibitem{Pratt_1992}
R.~G. Pratt and C.~H. Chapman,
\newblock ``Traveltime tomography in anisotropic media—ii. application,''
\newblock {\em Geophysical Journal International}, vol. 109, no. 1, pp. 20--37,
  1992.

\bibitem{madariaga_seismic_2007}
R~Madariaga,
\newblock ``Seismic source theory,''
\newblock in {\em Treatise on Geophysics}, G~Schubert, Ed., vol.~4, pp. 59--82.
  Elsevier, 2007.

\bibitem{Tropp2006SPa}
J.~A. Tropp, A.~C. Gilbert, and M.~J. Strauss,
\newblock ``Algorithms for simultaneous sparse approximation. part {II}:
  {C}onvex relaxation,''
\newblock {\em Signal Processing, special issue on Sparse approximations in
  signal and image processing}, vol. 86, pp. 572--588, April 2006.

\bibitem{Beck:2009gh}
Amir Beck and Marc Teboulle,
\newblock ``{A Fast Iterative Shrinkage-Thresholding Algorithm for Linear
  Inverse Problems},''
\newblock {\em SIAM Journal on Imaging Sciences}, vol. 2, no. 1, pp. 183--202,
  Jan. 2009.

\bibitem{Bertsekas:2011ww}
Dimitri~P Bertsekas,
\newblock ``{Incremental gradient, subgradient, and proximal methods for convex
  optimization: a survey},''
\newblock {\em Optimization for Machine Learning}, p.~85, 2011.

\end{thebibliography}

\end{document}